\documentclass{article}
\usepackage{spconf,amsmath,graphicx}
\usepackage{makecell}

\title{IMPROVEMENT OF AUDIOVISUAL QUALITY ESTIMATION USING A NONLINEAR AUTOREGRESSIVE EXOGENOUS NEURAL NETWORK AND BITSTREAM PARAMETERS}
\name{Koffi Kossi, Stéphane Coulombe, Christian Desrosiers and Ghyslain Gagnon }%\thanks{Thanks to XYZ agency for funding.}}
\address{ Department of Software and IT Engineering \\
Department of Electrical Engineering \\
École de technologie supérieure, Université du Québec \\
Montréal, Québec, Canada
}
\begin{document}
%\ninept
%
\maketitle
\begin{abstract}
With the increasing demand for audiovisual services, telecom service providers and application developers are compelled to ensure that their services provide the best possible user experience. Particularly, services such as videoconferencing are very sensitive to network conditions. Therefore, their performance should be monitored in real time in order to adjust parameters to any network perturbation. In this paper, we developed a parametric model for estimating the perceived audiovisual quality in videoconference services. Our model is developed with the nonlinear autoregressive exogenous (NARX) recurrent neural network and estimates the perceived quality in terms of mean opinion score (MOS). We validate our model using the publicly available INRS bitstream audiovisual quality dataset. This dataset contains bitstream parameters such as loss per frame, bit rate and video duration. We compare the proposed model against state-of-the-art methods based on machine learning and show our model to outperform these methods in terms of mean square error (MSE=0.150) and Pearson correlation coefficient (R=0.931).
\end{abstract}
\section{Introduction}
\label{sec:intro}

A recent study conducted by Cisco predicts that IP video traffic will account for 82\% of all Internet traffic by 2022. The sum of all forms of IP videos, which include video streaming, video files exchanged through file sharing, video-streamed gaming, and videoconferencing, will continue to grow in the range of 80 to 90\% of total IP traffic~\cite{Article_1}.\\
With this increasing demand for video services, audiovisual quality assessment becomes an important task to succeed in such competitive markets. The standard way of measuring perceived audiovisual quality is to conduct subjective tests where the end user is asked to assess the audio/video quality as a whole or in various parts on a continuous or discrete scale. However, assessing quality by conducting subjective tests is resource consuming and is not feasible in real-time communications.\\
When studying the characteristics of a real-time multimedia communication, it is necessary to have a clear understanding of the impact of the available quality of service (QoS) parameters on perceived audiovisual quality~\cite{Article_2}. The QoS handles purely the technical aspects of a service~\cite{Article_3} while the perceived quality incorporates human-related factors.\\
In this paper, we propose a nonlinear autoregressive exogenous (NARX)-based model, which uses QoS parameters to estimate the perceived audiovisual quality from for real-time audiovisual services such as videoconference. Specifically, we consider the persistence aspect of videoconferences and add the temporal dimension to the perceived audiovisual quality. We suppose that the feedback from recent events is important to estimate the current perceived audiovisual quality. In addition, we show that this feedback information increases the accuracy of the quality estimation model. This could encourage service providers to adopt deep learning algorithms to optimize and control their services.\\
To the best of our knowledge, the NARX algorithm has not yet been applied to estimate the perceived audiovisual quality. This is primarily due to the quality perception simplification that results from ignoring its temporal aspect. Another reason is the fact that most of existing video databases have not considered the sequential aspect of videos. In this paper, we experimentally show that the NARX model can outperform state-of-the-art machine learning algorithms for audiovisual quality estimation.

\section{Related work}
\label{sec:related-works}

A large body of work was conducted on video quality assessment based on QoS parameters, however, most of them were performed for video streaming services. In~\cite{Article_4}, the authors present a broad review of perceived quality modeling and methodologies. In the specific case of audiovisual services, standardized models have achieved good prediction performance. More recently, various machine learning methods, for instance based on deep learning, were proposed to estimate the perceived audiovisual quality. In this section, we briefly present some of these standardized methods developed for streaming services and video telephony applications. Afterwards, we summarize the recent works using machine learning to estimate the perceived audiovisual quality.\\
The ITU-T G.1070~\cite{Article_5} is a model recommended for video telephony service planning. This model is derived from the E-model~\cite{Article_6} which was standardized for telephony over IP. The model takes as inputs some video parameters such as codec type, packet loss rate, bit rate, key frame interval, frame rate, video format and display size. Its main goal is to estimate audiovisual quality, while taking into consideration human perception so that service planners can avoid under-engineering. The use of this model is limited to quality planning, and applications such as monitoring are not covered by the recommendation.\\
The ITU-T P.1201~\cite{Article_7} model is a non-intrusive model recommended for estimating the audiovisual quality of streaming services. This model provides separate estimation for audio, video, and audiovisual quality, where the output or quality estimation is rated on the five-point mean opinion score (MOS) scale. It considers parameters such as compression, packet loss, re-buffering, frame rate, video resolution and bit rate affect audiovisual quality. For mobile TV (lower-resolution application) and IPTV (higher-resolution application), the Root Mean Square Error (RMSE) and Pearson correlation values for audiovisual modeling were evaluated as 0.470 and 0.852, 0.435 and 0.911, respectively. However, by design, these methods have limited application areas and cover limited coding technologies such as IPTV. Therefore, with the rise of the deep learning algorithms, many recent studies have focused on extending and improving these models.\\
Goudarzi et al.~\cite{Article_8} used a regression model to estimate audiovisual quality using two parameters corresponding to packet loss rate and frame rate. Their work also investigated how network and application parameters influence the overall audiovisual quality. The proposed model was evaluated under 60 different test conditions, where videos were degraded with packet loss and using different bit rates, and achieved a Pearson correlation coefficient of 0.93 and RMSE of 0.237 for the audiovisual database used in experiments.\\
Recently, Demirbilek et al.~\cite{Article_9} extended the work of Goudarzi et al. for videoconferencing services using machine learning algorithms such as decision trees, random forest and multi-layer perceptron (MLP). In~\cite{Article_10}, the same authors updated the database of Gourdazi et al.~\cite{Article_8} with additional QoS parameters than packet loss and frame rate, and estimated the perceived audiovisual quality of different machine learning algorithms on this extended database. Futhermore, these two authors extend their audiovisual database with information from the stream (bitstream).\\
While autoregressive nonlinear networks like the NARX have led to impressive results in various applications~\cite{Article_11, Article_12}, their potential for videoconference audiovisual quality assessment based on quality of service (QoS) parameters (no-reference model) has so far not been investigated.

\section{PROPOSED NARX-BASED APPROACH}
\label{sec:propoed-model}

\subsection{NARX model}
The nonlinear autoregressive with exogenous inputs (NARX) is a class of recurrent neural network capable of efficiently modeling time series data with feedback connections~\cite{Article_13}. 
\begin{figure}[h]
\begin{minipage}[b]{1.0\linewidth}
  \centering
  \centerline{\includegraphics[width=8.5cm]{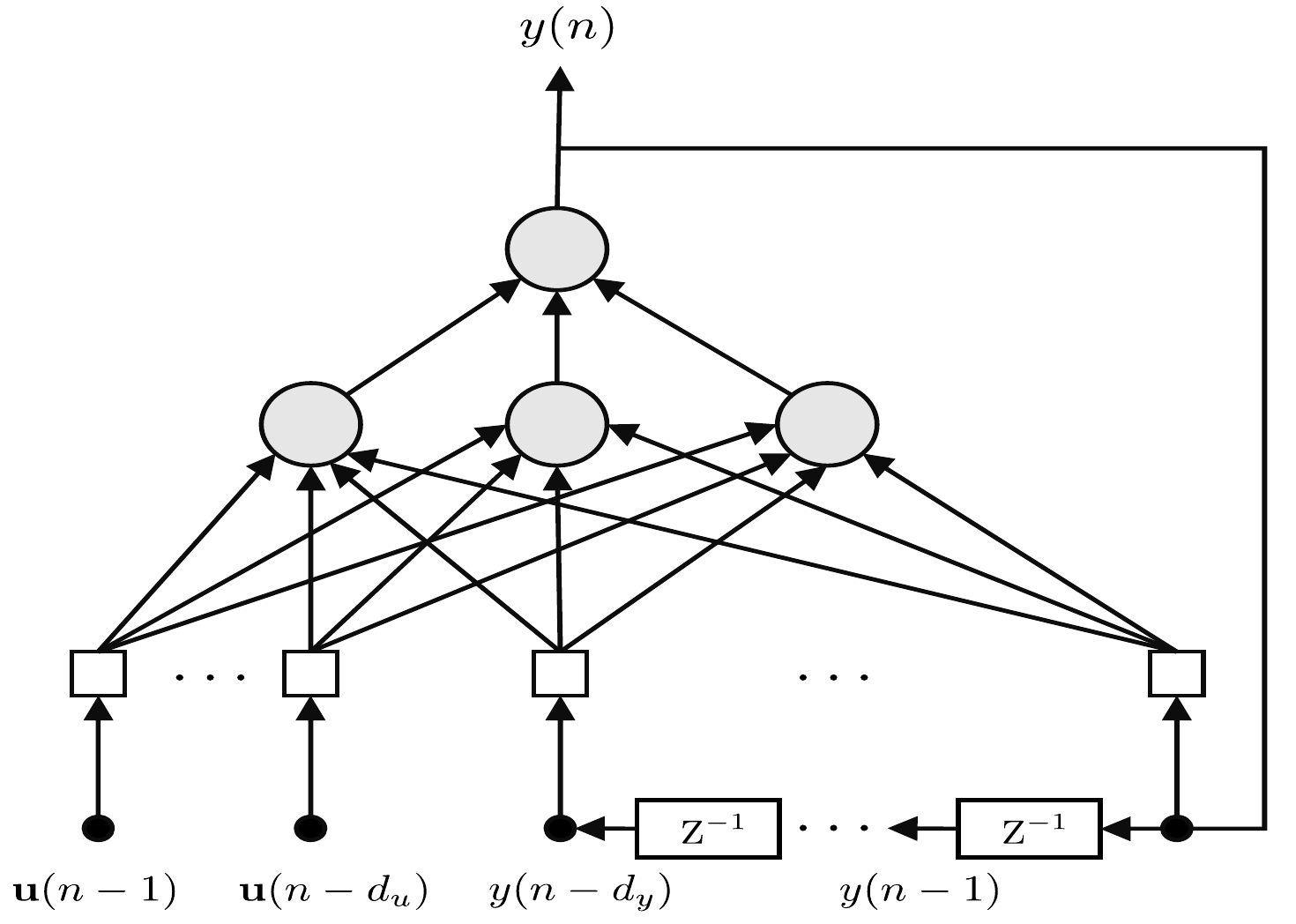}}
%  \vspace{2.0cm}
\end{minipage}
  \caption{\centering {NARX network with delayed inputs ($d_u$) and delayed outputs ($d_y$); $z^1$ is unit time delay.}}
  \label{fig:Model_architecture}
  %\medskip
\end{figure}

The NARX model has become popular in recent years because of its performance and fast convergence~\cite{Article_11, Article_12}. Some researchers even listing this model among the most efficient neural networks~\cite{Article_14}. The main difference between the NARX and conventional recurrent neural networks (RNNs) is that it uses feedback connections from the outputs to the inputs while RNNs have recurrent connections in the hidden layers~\cite{Article_15}. Siegelmann et al.~\cite{Article_14} have shown that gradient descent learning is more effective in NARX networks than in standard recurrent neural network architectures. When using the information from the exogenous inputs, the NARX model provides more flexibility compared with the feedforward network. This makes the model more effective than other neural networks and motivates its choice to solve our audiovisual quality estimation problem.\\
Studies have shown the robust performance of NARX for approximating almost every nonlinear function~\cite{Article_15}. The output $y(n)$ (Fig. 1) of the NARX model is the result of a nonlinear operation on multiple past outputs and an external variable:
\begin{equation}
\label{eq-NARX}
\begin{split}
y(n) \, = f\, ( \,y(n-1), \,y(n-2),\ldots.,\,y(n-d_{y});\\ \, \,u(n), \,u(n-1), 
\,\ldots.,\,u(n-d_{u}))
\end{split}
\end{equation}
From this equation, we see that there are two types of inputs to a NARX: past outputs $y(n-1)$, $y(n-2)$,  ... , $y(n-d_{y})$ that are fed back as future inputs (illustration of the dynamic model), and previous (and current) external variables $u(n)$, $u(n-1$), ... , $u(n-d_{u})$, where $d_{y}$ is the number of lags in the outputs and $d_{u}$ is the number of lags in the external variables (inputs). The autoregression function $f$ is approximated by a feed-forward multi-layer perceptron (MLP) neural network.\\
One should note that other neural networks can be derived from the NARX architecture. For example, without lag ($d_{y}=0$ and $d_{u}=0$), the model becomes a simple feed-forward multi-layer perceptron (MLP) network. Moreover, when there is no feedback loop of delayed outputs, the model is equivalent to a time delay neural network (TDNN)~\cite{Article_16}.
\subsection{Proposed NARX model to estimate the perceived audiovisual quality}
In this section, we propose a NARX-based model for estimating the perceived audiovisual quality using QoS parameters. As shown in Fig.\ref{fig:Model_architecture}, the basic NARX model consists of inputs corresponding to QoS-related parameters, a hidden layer and an output layer. 
In this work, we consider QoS parameters from the INRS bitstream audiovisual dataset such as aFrame count, pFrame count, video bit rate as inputs (see more details in Table 2). 
The output of the NARX network is the mean opinion score (MOS) ranging from 1 to 5. To reduce the model’s complexity, we use a single hidden layer with $N$ nodes. 
As recommended for the design of MLP networks~\cite{Article_15}, we set the number of neurons in the hidden layer equal to the total numbers of inputs parameters.\\ 
The hyperbolic tangent is used for the transfer function in the hidden layer, while a linear transfer function is employed for the output layer. For training, we selected the Bayesian regularization~\cite{Article_17, Article_18} training algorithm which reduces overfitting by penalizing the $L2$ norm of network parameters. During training, the inputs parameters and their corresponding MOS (from the INRS Bitstream dataset) are used to learn the model’s parameters. 
The inputs or QoS parameters such as packet loss are collected periodically in the videoconferencing gateway through RTCP (Real-Time Transport Control Protocol) reports. For example, the RFC 3550~\cite{Article_19} recommends the collection of these metrics every 5 seconds. Thus, the inputs data are sequential and could be considered as time-series data.\\
For the NARX model, in the presence of labeled data (supervised learning), the training method is called open loop (OL) configuration. During the testing, the estimated MOS value is obtained from the trained model by feeding it the inputs parameters.

\section{SIMULATION RESULTS AND ANALYSIS}
\label{sec:simulation}

In this section, we evaluate the performance of our proposed NARX-based model for perceived quality estimation and compare it against state-of-the-art machine learning methods for videoconferencing~\cite{Article_9}.

\subsection{INRS Bitstream Audiovisual Dataset}

Recently, INRS designed an audiovisual quality dataset namely “INRS Audiovisual Quality Dataset”~\cite{Article_9, Article_10}. This dataset was proposed to update the PLYM dataset~\cite{Article_8} which considered the importance of compression and network distortion in perceived quality estimation. 
Studies have shown that, for the videoconference service, the change in the transmission packet loss can have a dramatic impact on perceived quality. 
In addition, the video frame rate and quantization parameter may also affect quality, although more moderately.\\
Moreover, the authors of “INRS Audiovisual Quality Dataset” released two versions of this dataset called parametric and bitstream. 
While the parametric version is limited to 4 parameters namely packet loss rate, video quantization, noise reduction filters and video frame rate (see Table 1), the bitstream version collects information on group of frames for each video and contains 140 parameters retrieved in terms of $I$ and $P$ frames. These 140 parameters reported in the bitstream dataset are also referred as “Bitstream parameters”.

\begin{table}[htbp]
\begin{center}
\caption{ \centering{QoS parameters used to build INRS Audiovisual Dataset~\cite{Article_10}}}
\begin{tabular}{l|c}
\hline
\textbf{Parameters} &  \textbf{Values}\\
\hline
Frame per second (FPS) & 10, 15, 20, 25  \\
\hline
Quantization parameter (QP) & 23, 27, 31, 35  \\
\hline
Noise reduction filters (NRF) & 0.999 \\
\hline
Audiovisual packet loss rate (PLR) (\%) & 0, 0.1, 0.5, 1, 5 \\
\hline
\end{tabular}
\end{center}
\label{tab:INRS_Parameters}
\end{table}

Note that the two versions of “INRS Audiovisual Quality Dataset” are publicly available and for our study, we selected the bitstream version as the parameters in terms of $I$ and $P$ frames are relevant for the quality estimation. 
Furthermore, the bitstream version of the “INRS Audiovisual Quality Dataset” could be considered as the improvement of the parametric version because it contains both parametric values and information of headers and the payload of the video bitstream for each video. Thus, the bitstream version provides a maximum of information about the audio and video.\\ 
The video used to create the datasets were encoded with H.264 and a total of 160 degraded videos were obtained by applying various network conditions and application parameters to the reference or original video (see the values in Table 1). 
There is one $I$ frame every 10 seconds in the reference video and the total video has three $I$ frames. 
The subjective method used to obtain the MOS for each video is ACR (Absolute Rate Scaling) and the observers were allowed to provide their subjective scores after listening and watching to the first 10 seconds of the video~\cite{Article_9,Article_20}.\\
More detailed information about the INRS audiovisual quality dataset regarding the selected video sequence, test methodology and comparison to the other publicly available datasets is given in ~\cite{Article_9, Article_10}

As the bitstream dataset contains 140 parameters, we use random forest parameters features selection algorithm to obtain the top important parameters. From this operation, we got 9 important parameters as shown and defined in Table 2.
\begin{table}[htbp]
\begin{center}
\caption{ \centering{Nine (9) important parameters selected from INRS Bitstream Audiovisual Dataset ~\cite{Article_9}}}
\begin{tabular}{l|l}
\hline
\multicolumn{2}{c}{\textbf{Parameters}} \\
\hline
aFrame Count & Total count of audio frames   \\
\hline
pFrame Count Diff & \makecell[lt]{Total loss rate in the count of \\ pframes.}  \\
\hline
S2 pFrame Count Diff & \makecell[lt]{Loss rate in the count of P-frames  \\ for the 2$^{nd}$ second.} \\
\hline
S3 pFrame Mean Diff & \makecell[lt]{Loss rate in the average size of the \\ p-frames for the 10$^{th}$ second.} \\
\hline
S10 pFrame Mean Diff & \makecell[lt]{Loss rate in the average size of the \\ p-frames for the 10$^{th}$ second.}\\
\hline
\multicolumn{2}{c}{Audio NB Frames} \\
\hline
\multicolumn{2}{c}{Audio Bit Rate} \\
\hline
\multicolumn{2}{c}{Video Bit Rate} \\
\hline
\multicolumn{2}{c}{Video Packet Loss Rate} \\
\hline
\end{tabular}
\end{center}
\label{tab:INRS_Bitstream_Parameters}
\end{table}

To have a better estimation of performance, we employed a $k$-fold cross-validation strategy~\cite{Article_21}. In this strategy, the available data is split in $k$ even-sized subsets and, for each cross-validation fold, a different subset is set aside for testing while remaining examples are used to train the model. Training examples in each fold are further divided following a 70\%/30\% training-validation split for learning the network weights (training set) and selecting optimal training hyper-parameters (validation set). In this work, we used $k=5$ folds in our experiments.

\subsection{Results and Discussion}
\begin{figure}[h]
\begin{minipage}[b]{1.0\linewidth}
  \centering
  \centerline{\includegraphics[width=8.5cm]{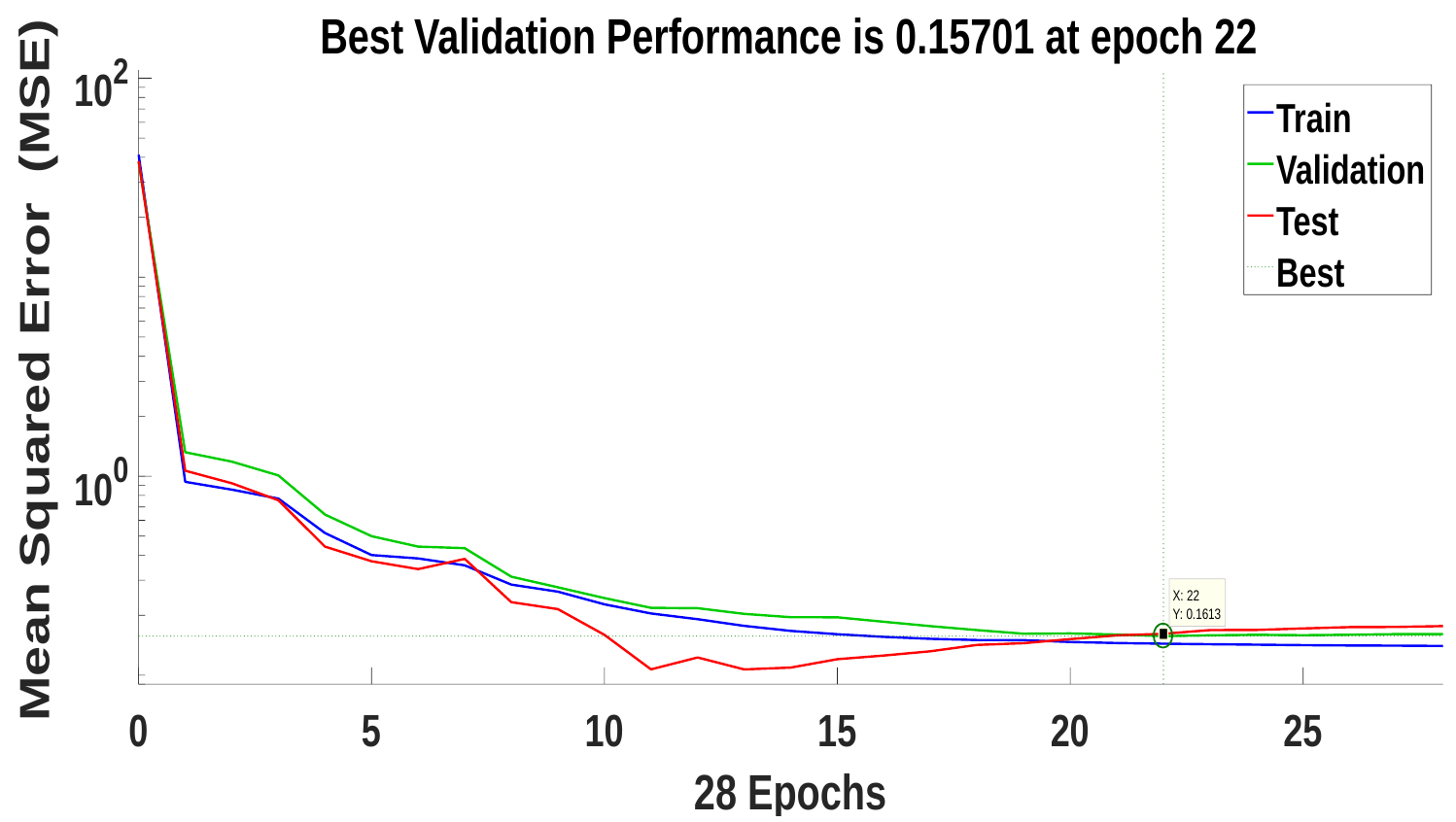}}
%  \vspace{2.0cm}
\end{minipage}
  \caption{\centering {Example of training, validation and test errors for a single fold.}}
  \label{fig:RMSE}
  %\medskip
\end{figure}

In the following experiments, the performance of our NARX-based model is compared to that of recent work on audiovisual services using the same database~\cite{Article_10}. This previous works showed that machine learning algorithms such as random forest (RF) perform better than the MLP algorithm for estimating the perceived audiovisual quality.\\
To evaluate the performance of our model under various configurations and compare it to the state-of-the-art algorithms, we chose two metrics recommended by the ITU-T P.1401~\cite{Article_23}: Pearson correlation coefficient (R) and Mean squared error (MSE). For selecting the model applied on test examples, we stopped the training when the validation error reached its minimum, as shown in Figure~\ref{fig:RMSE}.\\

Different topologies were tested for our NARX model, using either no feedback of delayed outputs ($d_y = 0$) or a feedback of delayed outputs equal to that of delayed inputs ($d_y = d_u$). In the latter case, the model estimates the perceived audiovisual quality (estimated MOS) at time $t$ based on the values of the QoS parameters (inputs) at time $t-1$, $t-2$, …,  $t-d_u$ and past outputs $y(t-1)$, $y(t-2)$, …, $y(t-d_y)$.

Table 3 reports the performance obtained for three settings of our NARX model:  no feedback of delayed inputs with $d_u = 3$, feedback of delayed inputs and outputs with $d_y = d_u = 3$, and feedback of delayed inputs and outputs with $d_y = d_u = 4$. Note that other settings were tested, however, they did not improve the performance; hence we do not report them in the tables. As in~\cite{Article_9}, we simulated and compared our results with the random forest, linear regression, bagging and MLP algorithms. 

\begin{table}[htbp]
\begin{center}
\caption{ \centering{Performance Comparison of Various Models on the INRS Bitstream Audiovisual Dataset~\cite{Article_10}}}
\begin{tabular}{l|c|c}
\hline
\textbf{Model} &  \textbf{MSE} &  \makecell{\textbf{Pearson} \\ \textbf{Correlation}}\\
\hline
Multi Layer Perceptron (MLP) & 0.270  & 0.851\\
\hline
Random Forest (RF) & 0.143 & 0.921\\
\hline
Bagging & 0.160 & 0.912 \\
\hline
 Our NARX ($d_y= 0, d_u=3$)&0.203 & 0.911 \\
\hline
Our NARX ($d_y= d_u=3$)& 0.144 & 0.932 \\
\hline
Our NARX ($d_y= d_u=4$)&0.170 & 0.931 \\
\hline
\end{tabular}
\end{center}
\label{tab:Performance_INRS_Bitstream_DB}
\end{table}

We observe that our proposed NARX-based perceived audiovisual quality estimation model outperforms the three other audiovisual quality models for all three tested settings of delayed inputs and outputs. Comparing the settings together, the best performance is achieved for input and output lags of $d_y = d_u = 3$, with Pearson correlation coefficient of 0.932 and MSE of 0.144 (see Table 3).  
When using the bitstream dataset, we obtain a clear understanding of these setting as the database is set with a reference video that has three $I$ frames. So, the input and output lags of $d_y = d_u = 3$ is the best setting for the videos in this dataset. 
Thus, the best performance is obtained when using the information about the three $I$ frames in the video during the quality prediction. When, we increase the input and output lags i.e  $d_y = d_u = 4$, there is no increasing in Pearson correlation and also we obtained the value of MSE increased. The others frames don’t add any improvement to the prediction. 
\begin{figure}[htbp]
\begin{minipage}[b]{1.0\linewidth}
  \centering
  \centerline{\includegraphics[width=8.5cm]{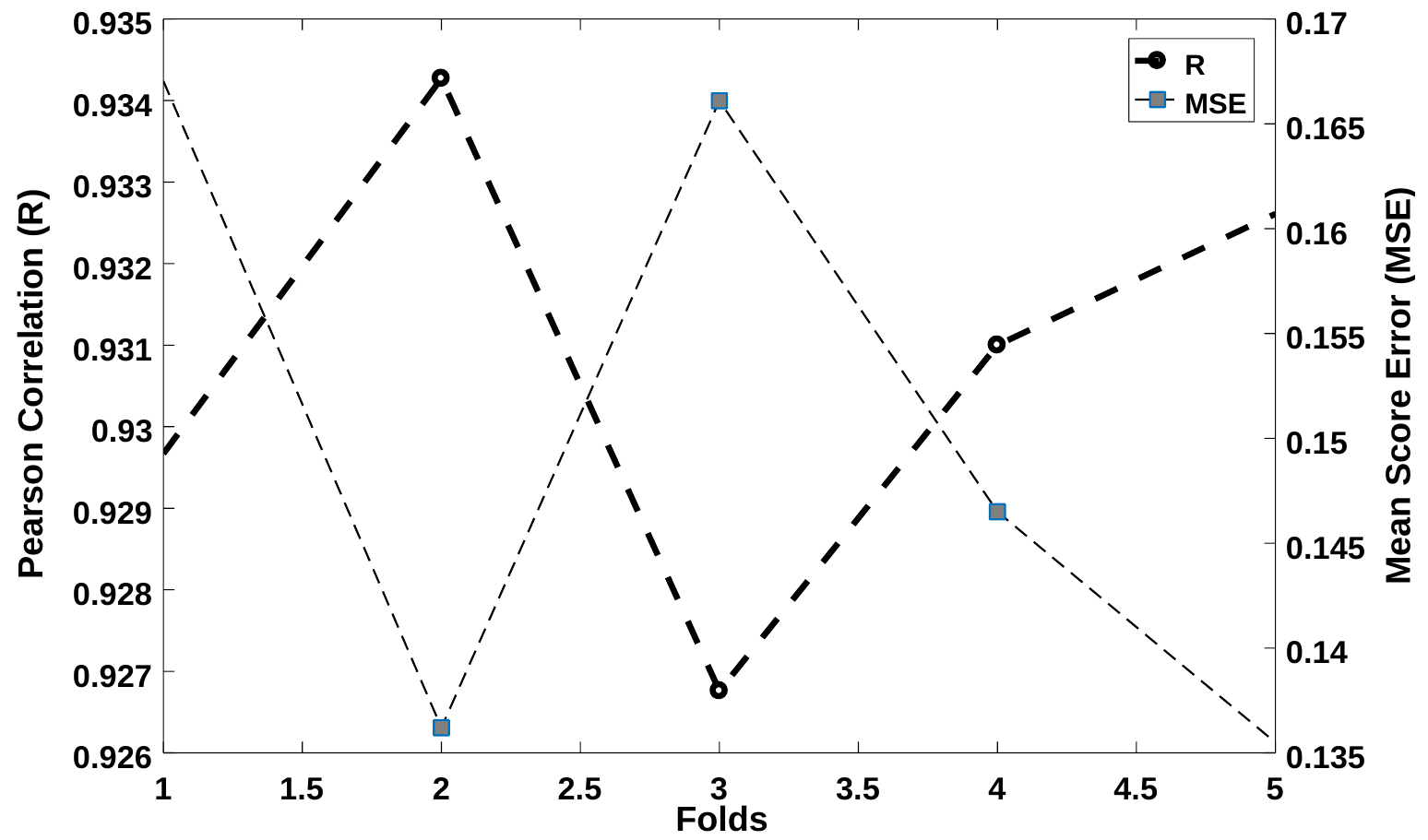}}
%  \vspace{2.0cm}
\end{minipage}
  \caption{\centering {Performance in terms of Pearson Correlation (R) and Mean Squared Error (MSE) for 5-fold}}
  \label{fig:Pearson_Correlation_R}
  %\medskip
\end{figure}

Results also show the usefulness of performing autoregression on both inputs and outputs, as highlighted by the improvement in MSE and Pearson correlation compared to the setting with $d_y = 0$ and $d_u = 3$. Increase the number of delayed inputs and outputs also leads to a higher performance, although this trend does not continue beyond a lag of 4. Figure~\ref{fig:Pearson_Correlation_R} gives the MSE and Pearson correlation values obtained for individual folds, while using $d_y = d_u = 3$. As can be seen, the performance remains stable across folds with a small variance in both metrics.

\section{CONCLUSION}
\label{sec:conclusion}

In this paper, we presented a no-reference NARX-based perceptual quality estimation model for audiovisual service. In the proposed model, the application and network parameters such as the frame rate, quantization parameter, packet loss rate and noise reduction filters were taken into consideration. We evaluated and compared our proposed NARX-based perceptual quality estimation model using the INRS audiovisual dataset. Simulation results showed that our proposed NARX-based model outperforms the state-of-the-art models in terms of Pearson correlation coefficient and MSE. The results indicate that our NARX-based perceived audiovisual quality estimation is a practical method for the assessment of video perceived quality. Moreover, with only a single hidden layer, our NARX-based model can well estimate the perceived quality. Future works will target other QoS parameters having an impact on perceived quality and use them to improve the results of the perceived audiovisual quality estimation.

\subsection{ACKNOWLEDGEMENT}
\label{ssec:subhead}

This work was supported in part by the Natural Sciences and Engineering Research Council of Canada (NSERC), and in part by Dr. Richard J. Marceau Chair on Wireless IP Technology for Developing Countries, particularly through our industrial partner, Media5 Corporation.

 %\section{RELATION TO PRIOR WORK}
 %\label{sec:prior}

 %\cite{Lamp86} considers only fixed time-domain analysis and the work by Jones et al 

\vfill\pagebreak

%\section{REFERENCES}
%\label{sec:refs}

\bibliographystyle{IEEEbib}
\bibliography{refs}

\end{document}